\title{Optimal Placement Algorithms for Virtual Machines}
\author{
        Umesh Bellur\\
        Department of Computer Science\\
        Indian Institute of Technology Bombay\\
        Mumbai 400076, India
	    \and
        Chetan S. Rao \\
        Department of Computer Science \& Engineering\\
        National Institute of Technology Calicut\\
        Kerala 673601, India
        \and
	    Madhu Kumar S.D.\\
        Department of Computer Science \& Engineering\\
        National Institute of Technology Calicut\\
        Kerala 673601, India
}
\date{\today}

\documentclass[11pt]{article}
\usepackage{amsmath}
\usepackage{algorithm}
\usepackage{algorithmic}
\usepackage{graphicx}
\usepackage{hyperref}

\newcounter{theorem}

\newcommand{\theoremlike}[1]{\par\medskip\penalty-250\refstepcounter{theorem}{\bfseries\scshape\noindent#1 \thetheorem.}\slshape}
\newenvironment{theorem}{\theoremlike{Theorem}}{\par\medskip}

{\par\medskip}
{\par\medskip}

\newcommand{\proofofthm}{{\scshape\noindent Proof.\quad}}
\newcommand{\qedofthm}{\hfill\rule{1ex}{1em}\penalty-1000{\par\medskip}}

\begin{document}
\maketitle
\begin{abstract}
Cloud computing provides a computing platform for the users to meet their demands in an efficient, cost-effective way. Virtualization technologies are used in the clouds to aid the efficient usage of hardware. Virtual machines (VMs) are utilized to satisfy the user needs and are placed on physical machines (PMs) of the cloud for effective usage of hardware resources and electricity in the cloud. Optimizing the number of PMs used helps in cutting down the power consumption by a substantial amount.

In this paper, we present an optimal technique to map virtual machines to physical machines (nodes) such that the number of required nodes is minimized. We provide two approaches based on linear programming and quadratic programming techniques that significantly improve over the existing theoretical bounds and efficiently solve the problem of virtual machine (VM) placement in data centers.
\end{abstract}

\section{Introduction}\label{ch:overview}

Cloud computing is a large scale network-based distributed computing environment where computing resources such as memory, processing power, bandwidth, etc. are available on demand to the users. The cloud computing environment comprises of many models such as \textit{Software as a Service (SaaS)}, \textit{Platform as a Service (PaaS)} and \textit{Infrastructure as a Service (IaaS)}. These models are made available to the users through virtualization techniques. The users' demands are satisfied by a set of servers hosted on virtual machines (VMs). The VMs utilize the resources of underlying physical machines (PMs) or nodes provided and operated by organizations called `cloud providers'. Some examples of cloud providers include Amazon EC2~\cite{LinkA}, GoGrid~\cite{LinkG} and Rackspace Cloud~\cite{LinkR}.

Adopting the use of virtual machines (VMs) in such large-scale environments enhances the number of available servers through multiple OS instances on a single node, thereby achieving efficient hardware utilization. However, there may be a number of underutilized nodes due to the inefficient mapping of virtual machines to physical machines. Minimizing the number of physical machines utilized helps in cutting down the power consumption drastically~\cite{BR}.

The placement algorithm for VMs in a data center allocates
various resources such as memory, bandwidth, processing power, etc. from a physical
machine (PM) to VMs such that the number of PMs used is minimized.

\begin{figure*}
\centering
\begin{tabular}{l l}
\includegraphics[scale=0.3]{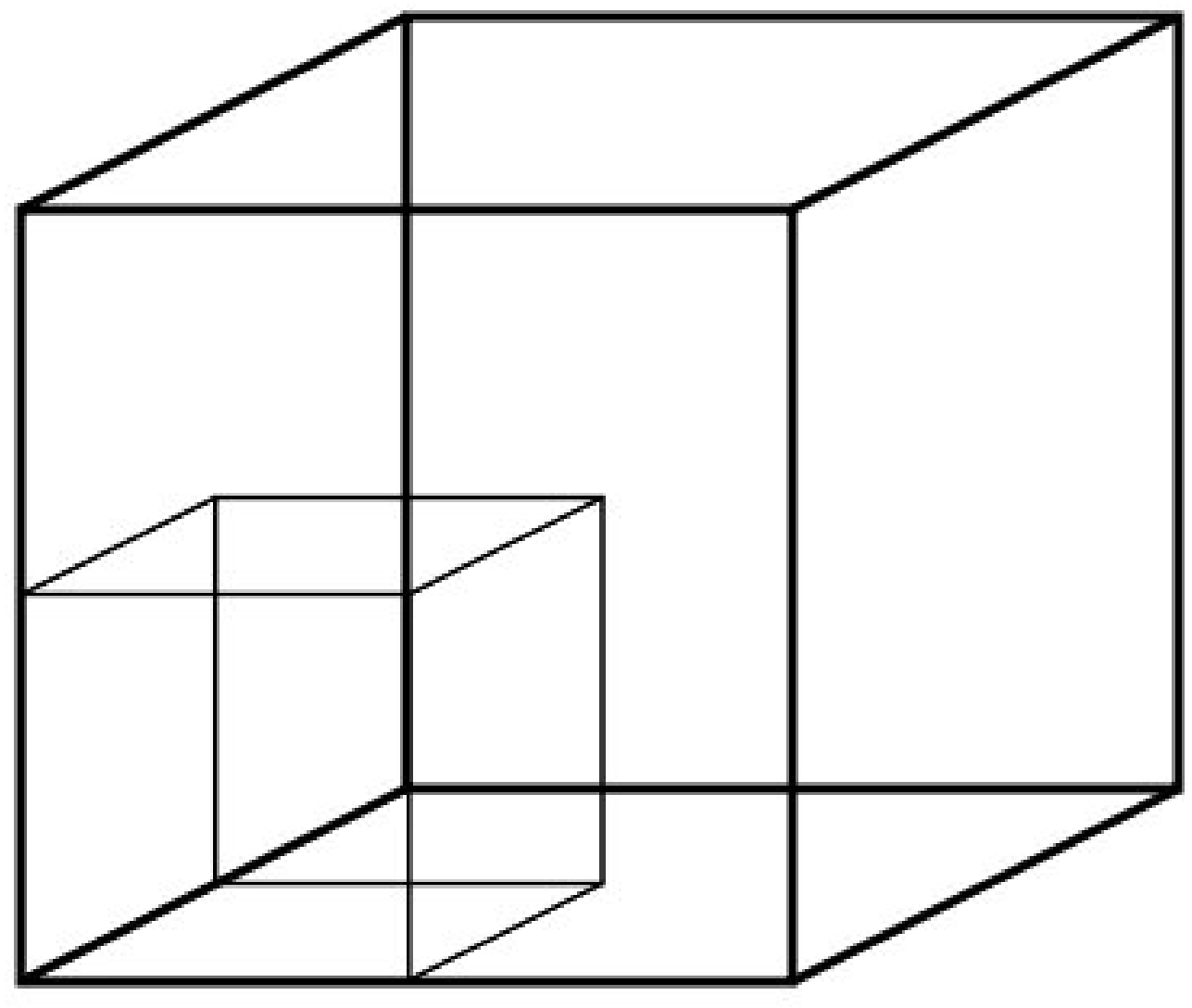} &
\includegraphics[scale=0.3]{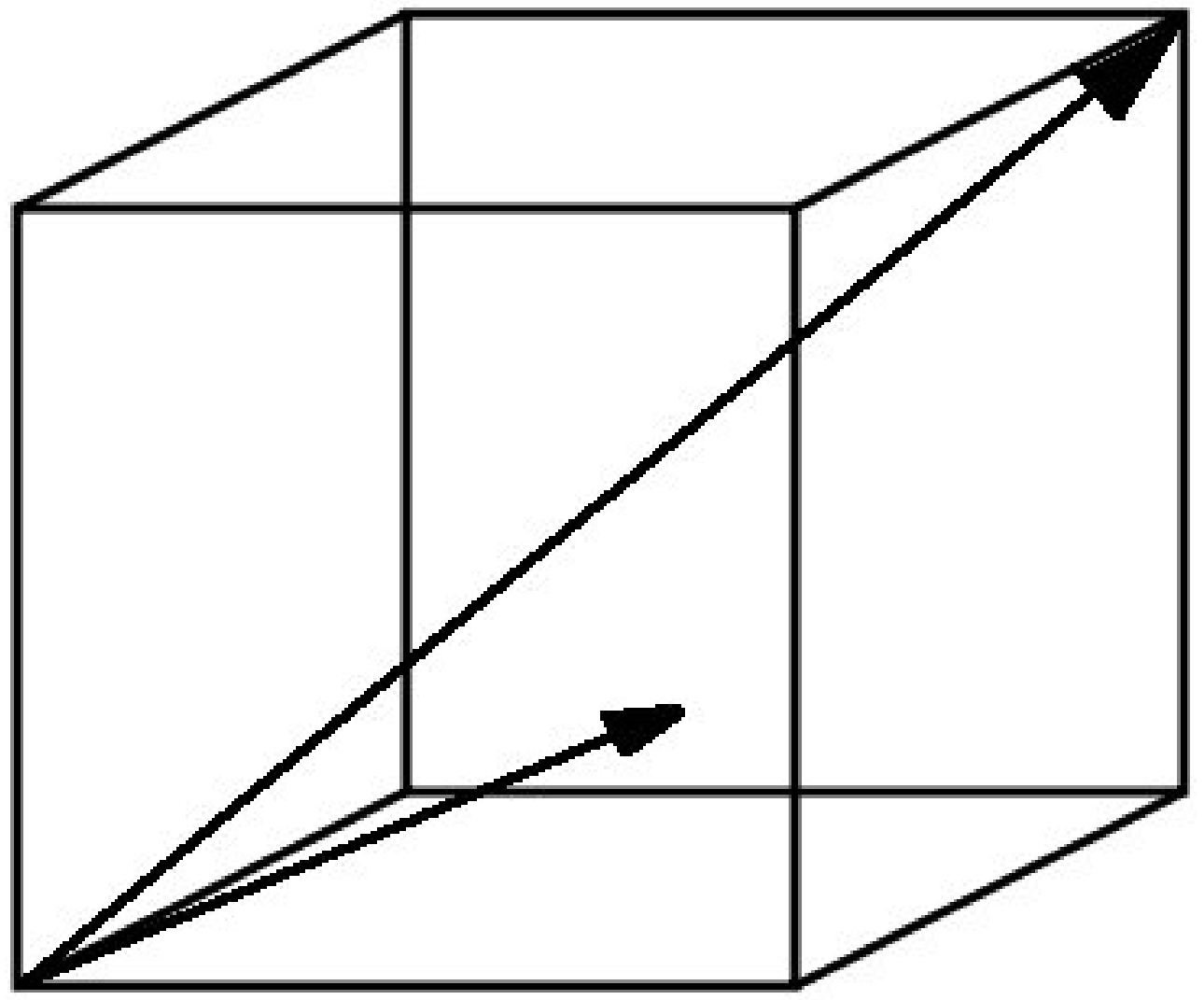} \\
\end{tabular}
\caption{Bin-packing \& Vector Bin Packing along 3-dimensions}\label{fig:vbp}
\end{figure*}

This problem can be viewed as a multi-dimensional packing problem~\cite{GGJ} (Figure \ref{fig:vbp}). The resource
requests of VMs are considered as $d$-dimensional vectors with non-negative entries (balls).
The resource available at each PM is considered to be a $d$-dimensional vector (each dimension signifies an independent resource) with a magnitude of 1 along each dimension (bins).
The goal is to minimize the number of bins such that for every bin the sum of the vectors
placed in that bin is coordinate-wise no greater than the bin's vector. Thus, the resource
allocation problem is an instance of the $d$-dimensional Vector Bin Packing problem (VBP)~\cite{GGJ}.

For d = 1, the VBP is identical to the 1-dimensional Bin Packing problem.

We now define the optimization problem that we are addressing in this paper.\\ \\
\textbf{Vector Bin Packing problem} (VBP)\\
\textit{Given a set S of `n' d-dimensional vectors $p_{1}$, $p_{2}$, \ldots, $p_{n}$ from [0,1]$^{d}$, find a packing (partition) of S into $A_{1}$, $A_{2}$, \ldots, $A_{m}$ such that $\sum_{p \in A_{i}} p^k \leq 1,$ $\forall i, k$ ($p^{k}$ denotes the projection of vector p along `k'th dimension). The objective is to minimize the value of `m', the number of partitions.}\\

The vector bin packing problem is a computationally hard problem and it is known to be NP-Hard~\cite{GJ}.
\\ \\

\section{Related work}

\vspace*{-1mm}
\paragraph{Vector Bin Packing (VBP)}
One dimensional bin packing problem has been studied extensively. Fernandez
de la Vega and Lueker~\cite{FL} gave the first Asymptotic Polynomial-Time
Approximation Scheme (APTAS). They put forward a rounding technique that 
allowed them to reduce the problem of packing large items to finding an optimum
packing of just a constant number of items (at a cost of $\epsilon$ times the optimal solution - $\mathsf{OPT}$).
Their algorithm was later improved by Karmarkar and Karp~\cite{KK}, to a 
(1+$log^{2}$)-$\mathsf{OPT}$ bound.

For 2-dimensional vector bin packing, Woeginger~\cite{W} proved that there is no
APTAS. For higher dimensions, Fernandez de la Vega and Lueker~\cite{FL} proposed a 
simple ($d + \epsilon$)-$\mathsf{OPT}$ algorithm, which extends the idea of 1-dimensional bin packing.
Chekuri and Khanna~\cite{CK} showed an $O$(log $d$)-approximation algorithm that
runs in polynomial time for fixed d. Bansal et al.~\cite{BCS} improved this result,
showing a (ln $d$ + 1 + $\epsilon$)-approximation algorithm for any $\epsilon \ge 0$. Karger et 
al.~\cite{KO} have recently proposed a polynomial approximation scheme for randomized instances of 
the multidimensional vector bin packing using \textit{smoothing techniques}. Patt-Shamir et al.~\cite{BPS} 
have recently explored the vector bin packing problem with bins of varying sizes and propose a 
(ln $2d$ + 1 + $\epsilon$)-approximation algorithm for any $\epsilon \ge 0$.

\vspace*{-1mm}
\paragraph{Placement Algorithm}
The problem of VM placement is at the core of cloud computing. Several research works address the importance of placing VMs appropriately~\cite{GIYC, BR, CKS}. Vogels~\cite{V} quotes the benefit of packing VMs efficiently in server consolidation. Recently, Hermenier et al.~\cite{HLM} developed a contraint programming based mechanism for dynamic consolidation.

Several modified versions of First-Fit Decrease (FFD) have been used for VM placements. Verma et al.~\cite{VAN} propose an algorithm to pack VMs optimally while minimizing the number of migrations. Khanna et al.~\cite{KBKK} propose a reconfiguration algorithm to cut down the wastage of physical resources. Hyser et al.~\cite{HMGW} propose an iterative rearragement technique for improving placements in a dynamic scenario. Bobroff et al.~\cite{BKB} presents a dynamic algorithm that forecasts the resource demands and packs VMs. Shahabuddin et al.~\cite{SCGJKK} propose a simple heuristic which aims to efficiently allocate resources.

Despite the recent research trends towards virtualization, the problem of VM placements is vastly unexplored. To overcome this limitation, we propose a linear programming based approach which places VMs efficiently on a set of PMs. 

The rest of this paper is organized as follows. Section \ref{probform} deals with the formulation of the problem, Section \ref{algo} provides our algorithms for vector bin packing (VBP) and Section \ref{expres} describes the experimental setup and results. Section \ref{conclusion} concludes the paper.

\section{Problem formulation} \label{probform}
We formulate the problem as an integer program in subsection \ref{ilpform}. The integer constraints are relaxed and we formulate
it's dual (in subsection \ref{dualform}). The solution of the relaxed integer linear program gives a thoughtful insight about the optimal number of bins.

\begin{figure}
\begin{center}
\begin{tabular}{|l|l|l|}
\hline
\multicolumn{2}{|c|}{\textbf{Notation Table}} \\
\hline
$x_{ij}$ & Fraction of vector $i$ packed in $j$th bin \\ \hline
$y_{j}$ & Binary variable to determine usage of bin $j$ \\ \hline
$p_{i}$ & Input vector $i$ (VM$_{i}$) \\ \hline
$n$ & Number of vectors (VMs) \\ \hline
$m$ & Number of bins (PMs) \\ \hline
$d$ & Dimension of each vector \\ \hline
\end{tabular}
\end{center}
\caption{Notation table for the integer linear program (ILP) formulation} \label{nottab}
\end{figure}

\subsection{Integer Linear Program (ILP) formulation} \label{ilpform}
The vector bin packing problem (\textsc{VBP}) can be formulated as an integer program. We use two binary variables $x_{ij}$ and $y_{j}$. The binary variable $x_{ij}$ indicates if vector $p_{i}$ is assigned to bin $j$ and the binary variable $y_{j}$ indicates whether bin $j$ is in use or not. Our objective is to minimize the number of bins used.

The number of bins $m$ can initially be set to a sufficiently large value arrived at by any heuristic (example - de la Vega and Leuker~\cite{FL} give a $O(d)$-$\mathsf{OPT}$ bound on the number of bins). Then, we formulate the integer program (\textsc{ILP}) as follows -

\begin{align}
    \mbox{minimize :\hspace{3mm}} \sum_{j} y_{j} \mbox{\hspace*{5mm}s.t.} \qquad \qquad \\
    \sum_{j} x_{ij} = 1  \qquad \qquad \qquad \qquad 1 \leq i \leq n \label{lp1} \\ 
    \sum_{i} p_{i}^{k} . x_{ij} \leq 1 \qquad 1 \leq j \leq m, 1 \leq k \leq d \label{lp2}\\
    y_{j} \geq x_{ij} \qquad \qquad 1 \leq i \leq n, 1 \leq j \leq m\\
    x_{ij} \in \mbox{\{}0, 1\mbox{\}} \quad \qquad 1 \leq i \leq n, 1 \leq j \leq m
\end{align}

The notations are mentioned in Figure \ref{nottab}. The constraints of the ILP are as follows -
\begin{itemize} 
\item Constraint (2) states that every vector is packed in a bin. 
\item Constraint (3) ensures that the packed vectors do not exceed the bin dimensions. 
\item Constraint (4) tells whether a bin is used or not.
\item Constraint (5) ensures that a vector is either packed entirely in a bin or not. 
\end{itemize}

Constraint (5) can be relaxed as follows to obtain a linear program (LP).
\begin{equation} \label{lp3}
	x_{ij} \geq 0 \quad \qquad 1 \leq i \leq n,\quad 1 \leq j \leq m  \tag{5a}
\end{equation}
We can obtain a feasible solution for the LP using any standard method~\cite{DA}. Using binary search technique, we can also find the least value of $m$, $m^{'} \in Z^{+}$ for the relaxed ILP for which a feasible solution exists. The value of $m^{'}$ thus obtained will be less than the optimal solution for the integer program i.e. ($m^{'} \leq $ $\mathsf{OPT}$). However, the solution obtained is usually not integral. To tackle this problem, we formulate a dual-maximization problem~\cite{VV} for the above relaxed ILP.

\subsection{Dual-maximization problem} \label{dualform}
We introduce several new variables  $z_{ij}$ to formulate the dual. The dual-maximization problem formulation is given in the Appendix \ref{appendix}. We arrive at the following set of equations and constraints -\\
\begin{align}
    \mbox{maximize :\hspace{3mm}} \sum_{i} \sum_{j} x_{ij}z_{ij} \qquad \mbox{s.t.} \qquad \\
    \sum_{j} x_{ij} = 1 \qquad \qquad \qquad \qquad 1 \leq i \leq n \\
    \sum_{i} p_{i}^{k} . x_{ij} \leq 1 \qquad 1 \leq j \leq m, 1 \leq k \leq d  \\
    \sum_{i} z_{ij} \leq 1 \qquad \qquad \qquad \qquad 1 \leq j \leq m \label{cons9} \\
    x_{ij}, z_{ij} \geq 0 \quad \qquad 1 \leq i \leq n, 1 \leq j \leq m
\end{align}

This is a nonlinear program (NLP) as the objective function is nonlinear. Hereafter, we shall refer to it as NLP.

The number of variables in the NLP can be reduced by performing the following substitutions for the value of $z_{ij}$'s. 
	
\begin{theorem} \label{th:5}
The optimal solution to the NLP will still be optimal when the value of $z_{ij}$ is replaced by $x_{ij}/\sum_{i} x_{ij}$.
\end{theorem}
\proofofthm From the Jensen's Inequality, we have that if $f$ is a convex function (``concave-up'') on an interval $I$ and $a_{i} \in I$ then for weights $\lambda_{i}$ summing to 1 -
\begin{align*}
	f(\sum_{i=1}^{n} \lambda_{i} a_{i}) \le \sum_{i=1}^{n} \lambda_{i} f(a_{i})
\end{align*}
We can apply Jensen's inequality with $\lambda_{i}$ and $a_{i}$ corresponding to $x_{ij}$ and $z_{ij}$, respectively. The modified set of equations in this case is as follows-  
\begin{align*}
	f(\sum_{j=1}^{m} x_{ij} z_{ij}) \le \sum_{j=1}^{m} x_{ij} f(z_{ij}) \quad \quad \quad 1 \le i \le n \tag{a} \label{eq:a}
\end{align*}
From the property of convex functions, we have -
\begin{align*}
	f(tx) \le tf(x) \quad \quad \quad 0 \le t \le 1 \tag{b} \label{eq:b}
\end{align*}
From (\ref{eq:a}) and (\ref{eq:b}), we have -
\begin{align*}
	f(\frac{1}{n} \sum_{i=1}^{n} \sum_{j=1}^{m} x_{ij} z_{ij}) \le \frac{1}{n} \sum_{i=1}^{n} \sum_{j=1}^{m} x_{ij} f(z_{ij}) \tag{c} \label{eq:c}
\end{align*}
Since $f(x)$ is a convex function, any value which maximizes $x$ also maximizes $f(x)$ and vice-versa. Hence, from the inequality (\ref{eq:c}), we have that the term $x_{ij} f(z_{ij})$ should be maximized for the objective function to be maximized. Indirectly, $z_{ij}$ has to be maximized relative to the values of $x_{ij}$. The value of $z_{ij}$ is constrained by the constraint (\ref{cons9}), and hence we come up with the following tight function for $z_{ij}$ -
\begin{align}
	z_{ij} &= \frac{x_{ij}}{\sum_{i} x_{ij}} \label{eq:11} \\
	\implies \sum_{i} z_{ij} &= \sum_{i} \frac{x_{ij}}{\sum_{i} x_{ij}} = 1 \nonumber 
\end{align} \qedofthm

Since $\sum_{i} \sum_{j} x_{ij}=n$, the value of $\sum_{i} x_{ij} \approx n/m$ for an appropriately chosen value $m$. Thus, the objective function can be reduced to a quadratic term.

From (\ref{eq:11}), the NLP now becomes -
\begin{align}
    \mbox{maximize :\hspace{3mm}} \sum_{i} \sum_{j} x_{ij}^2 \qquad \mbox{s.t.} \qquad \label{NLP} \\
    \sum_{j} x_{ij} = 1 \qquad \qquad \qquad \qquad 1 \leq i \leq n \\
    \sum_{i} p_{i}^{k} . x_{ij} \leq 1 \qquad 1 \leq j \leq m, 1 \leq k \leq d  \\
    x_{ij} \geq 0 \quad \qquad 1 \leq i \leq n, 1 \leq j \leq m
\end{align}

The optimal solution to the above modified NLP - NLP$'$ - must be necessarily integral (follows from Theorem \ref{th:5}). In this light, we now present our algorithms which will provide the (near-)optimal integer solution.

\section{Algorithms and their \\ complexity} \label{algo}

In this section, we provide two algorithms to solve the vector bin packing (VBP) problem. The main idea is to harness the polynomial-time solvability of linear and quadratic programming techniques.

\subsection{Quadratic programming}
The quadratic program NLP$'$ can be solved using various efficient techniques such as interior points, active set~\cite{M}, gradient techniques or through the extensions of simplex algorithm~\cite{M}.

\paragraph{Complexity}
Kozlov et al.~\cite{KTK} presented a polynomial time algorithm for solving convex quadratic programs. Since our objective function is a convex function, NLP$'$ can be solved in polynomial time.

\subsection{Linear programming}
The relaxed version of the integer linear program (ILP) can also be used to derive (near-)optimal solutions for the VBP problem. The algorithm is as follows -

\algsetup{indent=2em}
\begin{algorithm}[h!]
  \caption{$\ensuremath{\mbox{\sc PackingVectors}}(P_{n}, d)$}\label{alg:packvectors}
  \begin{algorithmic}[1] \label{alg1}
    \REQUIRE A set of vectors $p_{1}, p_{2}, \ldots, p_{n}$; $P_{n}$.\\ Dimension of vectors $d$
    \medskip
    \STATE $(m,X) = SolveLP(P_{n}, d)$
    \IF {$m \ge \frac{n}{2}$}
       \RETURN $FirstFit(P_{n}, d)$
    \ELSIF {$m \le \sqrt{\frac{n}{d}}$}
       \RETURN $GreedyLP(P_{n},X,d)$
    \ELSE
	\RETURN $IterativePack(P_{n},X,d)$
    \ENDIF
  \end{algorithmic}
\end{algorithm}

Algorithm \ref{alg1} is an iterative algorithm which packs vectors in every iteration until the input is exhausted. The algorithm branches into 3 cases depending upon the solution returned by the relaxed integer linear program - LP.

\begin{algorithm}[h!]
  \caption{$\ensuremath{\mbox{\sc GreedyLP}}(P_{n},X,d)$} \label{alg:greedylp}
  \begin{algorithmic}[1] \label{alg2}
    \REQUIRE A set of vectors $p_{1}, p_{2}, \ldots, p_{n}$; $P_{n}$ and a set of $x_{ij}$ values $X$.
    \medskip
    \STATE $X^{'} = SortDescending(X)$
    \WHILE {$X^{'} \ne \Phi$}
    \STATE Remove the top element $x_{ij}$ in $X^{'}$
    \IF {vector $p_{i}$ fits in bin $j$}
	\STATE $Pack(i,j)$
	\STATE Remove $p_{i}$ from $P_{n}$
    \ENDIF
    \ENDWHILE
    \STATE $PackingVectors(P_{n}, d)$
  \end{algorithmic}
\end{algorithm}

Algorithm \ref{alg2} is a subroutine of Algorithm \ref{alg1} which packs the vectors greedily, given the solution set $X = \{x_{ij} \forall i,j\}$

\begin{algorithm}[h!]
  \caption{$\ensuremath{\mbox{\sc IterativePack}}(P_{n},X,d)$}\label{alg:iterativepack}
  \begin{algorithmic}[1] \label{alg3}
    \REQUIRE A set of vectors $p_{1}, p_{2}, \ldots, p_{n}$; $P_{n}$ and a set of $x_{ij}$ values $X$.
    \medskip
    \STATE $P_{n}^{'} = P_{n}$
    \STATE $Z = FindDualObj(X, d)$
    \FOR{$j=1$ \TO $m$}
	\IF{$ \sum_{i} x_{ij} z_{ij} \ge \frac{1}{2} $}
		\STATE $X_{j}^{'} = SortDescending(X_{j})$
		\STATE $X_{j}^{''} = RemoveLessThanHalf(X_{j}^{'})$
		\STATE $Pack(X_{j}^{''})$
		\STATE $P_{n}^{'} = P_{n}^{'} \backslash PackedVectors$
	\ENDIF
    \ENDFOR
    \STATE $PackingVectors(P_{n}^{'}, d)$
  \end{algorithmic}
\end{algorithm}

Algorithm \ref{alg3} is a subroutine of Algorithm \ref{alg1} which packs the bins having utility factor ($\sum_{i} x_{ij} z_{ij}$) more than half. It ensures that the efficiently assigned vectors are packed into their corresponding bins.

\paragraph{Complexity}
The subroutines of Algorithm \ref{alg1} - Algorithms \ref{alg2},\ref{alg3} - run in polynomial time. Solving the relaxed integer linear program can be done in polynomial time~\cite{AS}. The First Fit heuristic also runs in polynomial time. Thus, Algorithm \ref{alg1} runs in polynomial time.  

\section{Experimental setup and results} \label{expres}
We test our `\textsc{PackingVectors}' algorithm (Algorithm \ref{alg1}) discussed above with the existing theoretical worst-case bound for the vector bin packing (VBP) problem.

\paragraph{Tools used} 
The Mixed Integer Linear Programming (MILP) solver `\verb=lp-solve='~\cite{LPSolve} was used to derive exact solutions for randomized input instances (20 VM configurations). `\verb=lp-solve=' was also used as a linear program solver in Algorithm \ref{alg1}.

2000 iterations of randomized test inputs were performed for each dimension ranging from 2 to 10 ($2\leq d \leq 10$). The number of input VMs, $n$, were about 20 in each iteration.

\begin{figure}
\centering
\includegraphics[scale=0.5]{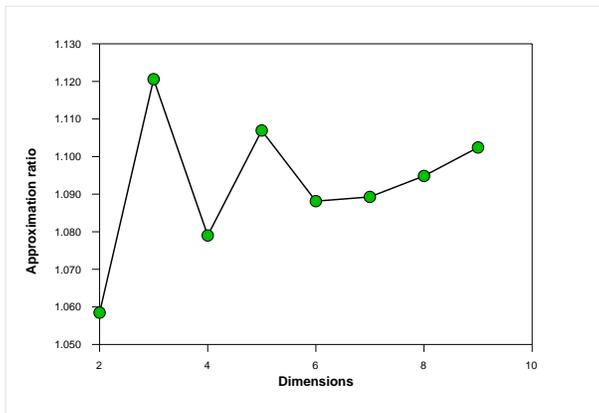} 
\caption{Mean approximation ratio - $\textsc{PackingVectors}$/$\mathsf{OPT}$. Mean ratio of about 2000 randomized trials along each dimension. For dimensions $d < 10$, the mean approximation ratio stays below 1.2}\label{stat1}
\end{figure}

Our results were compared with the exact solution of the optimal number of PMs (bins), and the mean approximation factor was computed. The mean approximation ratios are as shown in Figure \ref{stat1}.

\begin{figure}
\centering
\includegraphics[scale=0.5]{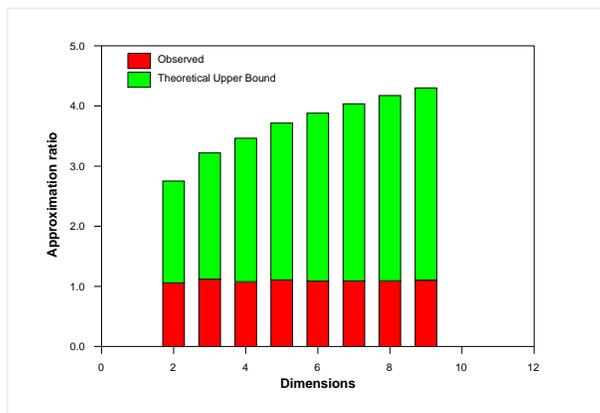} 
\caption{Our result vs. theoretical upper bound (ln $d$). The bars colored red indicate the mean approximation ratios of our algorithm whereas the green bars indicate the performance of the current best algorithm~\cite{BCS}} \label{stat2}
\end{figure}

Our results were also compared with the existing bounds of approximation given by Bansal et al.~\cite{BCS} and was found to have a substantial improvement as seen in Figure \ref{stat2}.

\section{Conclusions and future work} \label{conclusion}
We presented two novel algorithms for placement of VMs in data centers.
Unlike existing research based on simple First Fit heuristics, our techniques take advantage of the polynomial-time linear and quadratic programs, and provide (near-)optimal solutions to the vector bin packing (VBP) problem.

Our experiments confirm the substantial improvement of our approach over the existing techniques and demonstrate that our algorithm `\textsc{PackingVectors}' consistently yields the optimal placement across a broad spectrum of inputs. As part of future work, we intend to expand our techniques to handle dynamic placements and continuous optimization of data centers.

\section{Acknowledgments}
We would like to thank our colleagues 
for their help in the implementation of the `\textsc{PackingVectors}' algorithm.

\vspace*{10mm}

\section{Appendix}

\subsection{Dual formulation of the ILP} \label{appendix}
\begin{align} 
    & \mbox{minimize :\hspace{3mm}} \sum_{j} y_{j} & \\
    \mbox{such that \hspace*{3mm}} & \sum_{j} x_{ij} = 1 & 1 \leq i \leq n \label{1lp1}\\ 
    & \sum_{i} p_{i}^{k} . x_{ij} \leq 1 & 1 \leq j \leq m, 1 \leq k \leq d \label{1lp2}\\
    & y_{j} \geq x_{ij} & 1 \leq i \leq n, 1 \leq j \leq m \label{neq:4}\\
    & x_{ij} \in \mbox{\{}0, 1\mbox{\}} & 1 \leq i \leq n, 1 \leq j \leq m \label{1intgr}
\end{align}
Multiply constraint (\ref{neq:4}) by positive multipliers $z_{ij}$ corresponding to $x_{ij}$'s. Adding all such constraints, we obtain -
\begin{align*}
	(\sum_{i} z_{ij}) y_{j} &\ge \sum_{i} x_{ij} z_{ij} &\quad \quad \quad 1 \leq j \leq m \\
	\sum_{j} (\sum_{i} z_{ij}) y_{j} &\ge \sum_{j} \sum_{i} x_{ij} z_{ij} &
\end{align*}
Further, we have -
\begin{align}
	\sum_{j} y_{j} \ge \sum_{j} (\sum_{i} z_{ij}) y_{j} \ge \sum_{j} \sum_{i} x_{ij} z_{ij} \nonumber\\
	\mbox{ subject to \hspace*{5mm} } \sum_{i} z_{ij} \le 1 \quad \quad \label{1newcons}
\end{align}
Thus, the minimization problem can be reframed as a maximization problem with the constaint (\ref{1newcons}) and objective function being -
\begin{align*}
	\mbox{max : \hspace*{3mm}} \sum_{i} \sum_{j} x_{ij} z_{ij} &
\end{align*}
Adding the new constraints and relaxing constraint (\ref{1intgr}), the dual problem is as follows -
\begin{align*}
    & \mbox{maximize :\hspace{3mm}} \sum_{i} \sum_{j} x_{ij}z_{ij} & \\
    \mbox{such that \hspace*{3mm}} & \sum_{j} x_{ij} = 1 & 1 \leq i \leq n\\
    & \sum_{i} p_{i}^{k} . x_{ij} \leq 1 & 1 \leq j \leq m, 1 \leq k \leq d\\
    & \sum_{i} z_{ij} \leq 1 & 1 \leq j \leq m \\
    & x_{ij}, z_{ij} \geq 0 & 1 \leq i \leq n, 1 \leq j \leq m
\end{align*}
\end{document}